\documentclass[twocolumn,aps,prl,superscriptaddress,preprintnumbers,bibnotes]{revtex4-1}

\usepackage[T1]{fontenc}

\usepackage{natbib}
\usepackage{graphicx}
\usepackage{bm}
\usepackage{color}
\usepackage{amsmath}
\usepackage{gensymb} 
\usepackage{physics}
\usepackage{tabularx}
\usepackage{hyperref}

\newcommand\unit[2]{\ensuremath{#1~\mathrm{{#2}}}}

\newcommand{\SLJ}[3]{{\ensuremath{{^{#1}}\mathrm{#2}_{#3}}}}

\newcommand{\TPO}{\SLJ{3}{P}{1}}
\newcommand{\TPZ}{\SLJ{3}{P}{0}}
\newcommand{\TPJ}{\SLJ{3}{P}{J}}
\newcommand{\TSO}{\SLJ{3}{S}{1}}
\newcommand{\SSZ}{\SLJ{1}{S}{0}}
\newcommand{\SPO}{\SLJ{1}{P}{1}}
\newcommand{\TDO}{\SLJ{3}{D}{1}}

\newcommand{\Isotope}[2]{{\ensuremath{^{{#1}}\text{{#2}}}}}
\newcommand{\Sr}[1]{\Isotope{{#1}}{Sr}}

\begin{document}

\title{State-dependent optical lattices for the strontium optical qubit}

\author{A. Heinz}
\thanks{A.\,H. and A.\,J.\,P. contributed equally to this work.}
\author{A. J. Park}
\thanks{A.\,H. and A.\,J.\,P. contributed equally to this work.}
\author{N. \v{S}anti\'c}
\author{J. Trautmann}
\affiliation{
  Max-Planck-Institut f{\"u}r Quantenoptik,
  Hans-Kopfermann-Stra{\ss}e 1,
  85748 Garching, Germany}
\affiliation{
  Munich Center for Quantum Science and Technology,
  80799 M{\"u}nchen, Germany}
\author{S. G. Porsev}
\affiliation{Department of Physics and Astronomy, University of Delaware, Newark, Delaware 19716, USA}
\affiliation{Petersburg Nuclear Physics Institute of NRC ``Kurchatov Institute,'' Gatchina, Leningrad district 188300, Russia}
\author{M. S. Safronova}
\affiliation{Department of Physics and Astronomy, University of Delaware, Newark, Delaware 19716, USA}
\affiliation{Joint Quantum Institute, National Institute of Standards and Technology and the University of Maryland, College Park, Maryland, 20742, USA}
\author{I. Bloch}
\affiliation{
  Max-Planck-Institut f{\"u}r Quantenoptik,
  Hans-Kopfermann-Stra{\ss}e 1,
  85748 Garching, Germany}
\affiliation{
  Munich Center for Quantum Science and Technology,
  80799 M{\"u}nchen, Germany}
\affiliation{
  Fakult{\"a}t f{\"u}r Physik,
  Ludwig-Maximilians-Universit{\"a}t M{\"u}nchen,
  80799 M{\"u}nchen, Germany}
\author{S. Blatt}
\email{sebastian.blatt@mpq.mpg.de}
\affiliation{
  Max-Planck-Institut f{\"u}r Quantenoptik,
  Hans-Kopfermann-Stra{\ss}e 1,
  85748 Garching, Germany}
\affiliation{
  Munich Center for Quantum Science and Technology,
  80799 M{\"u}nchen, Germany}

\date{\today}

\begin{abstract}
  We demonstrate state-dependent optical lattices for the Sr optical qubit at the tune-out wavelength for its ground state. We tightly trap excited state atoms while suppressing the effect of the lattice on ground state atoms by more than four orders of magnitude. This highly independent control over the qubit states removes inelastic excited state collisions as the main obstacle for quantum simulation and computation schemes based on the Sr optical qubit. Our results also reveal large discrepancies in the atomic data used to calibrate the largest systematic effect of Sr optical lattice clocks.
\end{abstract}

\maketitle

\begin{figure}
  \centering
  \includegraphics{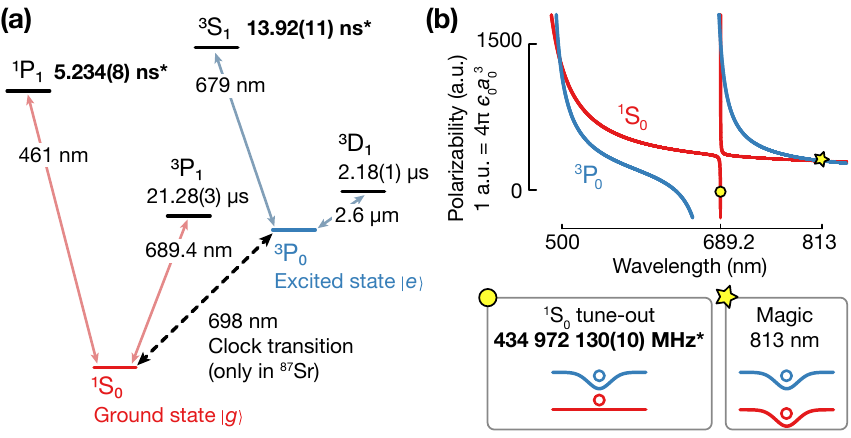}
  \caption{(color online). (a) Simplified Sr level diagram. Optical dipole traps in the red and infrared for the ground ($g$) and excited ($e$) qubit (clock) states of Sr are given by coupling to two low-lying singlet and triplet states, respectively. (b) The trap depth for each clock state as a function of wavelength is proportional to the dynamic dipole polarizability. At the \emph{magic wavelength} (star), $g$ and $e$ experience the same trap depth. At the \emph{tune-out wavelength} (circle), an atom in $g$ is free to move, while an atom in $e$ remains trapped. All quantities marked with an asterisk are measured in this work.}
  \label{fig:strontium}
\end{figure}

The experimental implementation of innovative quantum simulation and quantum computation schemes based on the optical qubit in strontium~\cite{daley08,gorshkov09,gorshkov10,fossfeig10a,fossfeig10b,daley11,daley11b} has been hindered by the presence of strong inelastic collisions between atoms in the excited qubit state~\cite{bishof11}.
Although these losses can be suppressed in deep three-dimensional optical lattices~\cite{campbell17},
such strong trapping precludes using tunneling and elastic collisions~\cite{martin13b} to entangle atoms in different lattice sites.
Controlled collisional phase gates~\cite{daley11b} in particular require high-fidelity, independent control over atoms in either qubit state \SSZ{} ($g$) and \TPZ{} ($e$),  shown in Fig.~\ref{fig:strontium}(a).
Here, we provide a solution to these problems by demonstrating optical lattices at the so-called \emph{tune-out wavelength} for the ground state~\cite{cheng13,safronova15}, where its dipole polarizability vanishes, as shown in Fig.~\ref{fig:strontium}(b).
At this tune-out wavelength, an $e$ atom is tightly trapped, while a $g$ atom is free to move.
This condition shuts off the inelastic $e$-$e$ collisions~\cite{bishof11}, while allowing the use of the elastic $e$-$g$ and $g$-$g$ collisions~\cite{martin13b} to engineer novel systems for quantum simulation~\cite{devega08,krinner18,tudela18,arguello19} and computation~\cite{daley11b}.

With a novel method, we measure an absolute frequency of \unit{434,972,130(10)}{MHz} for the tune-out wavelength in \Sr{88}.
At the tune-out wavelength, the differential AC Stark shift on the optical qubit transition is only due to the polarizability of the $e$ state.
We directly measure this polarizability with Stark shift spectroscopy, demonstrate trapping of $e$ atoms in an optical lattice at the tune-out wavelength, and show that losses from light scattering are small.
Given a moderate laser frequency stability corresponding to our measurement uncertainty, $e$ atoms are tightly trapped while the trap's effect on $g$ atoms is suppressed by more than four orders of magnitude, the highest level of suppression in any system to date~\cite{lamporesi10,holmgren12,trubko17,copenhaver19,herold12,leonard15,schmidt16,kao17,henson15}.

We combine these measurements with high-precision atomic structure theory and direct lifetime measurements of the \TPO{} state~\cite{nicholson15} to extract new values for the \SPO{} lifetime.
We find a 7$\sigma$ discrepancy to the currently accepted \SPO{} lifetime from photoassociative spectroscopy~\cite{yasuda06}.
Our polarizability measurements also improve the \TSO{} lifetime by an order of magnitude and resolve discrepancies between prior measurements~\cite{brinkmann69,havey77,jonsson84}.
Our results highlight the importance of direct and precise atomic lifetime measurements to bring the accuracy of optical lattice clocks~\cite{ludlow15} into the $10^{-19}$ regime.

\begin{figure*}
	\centering
	\includegraphics{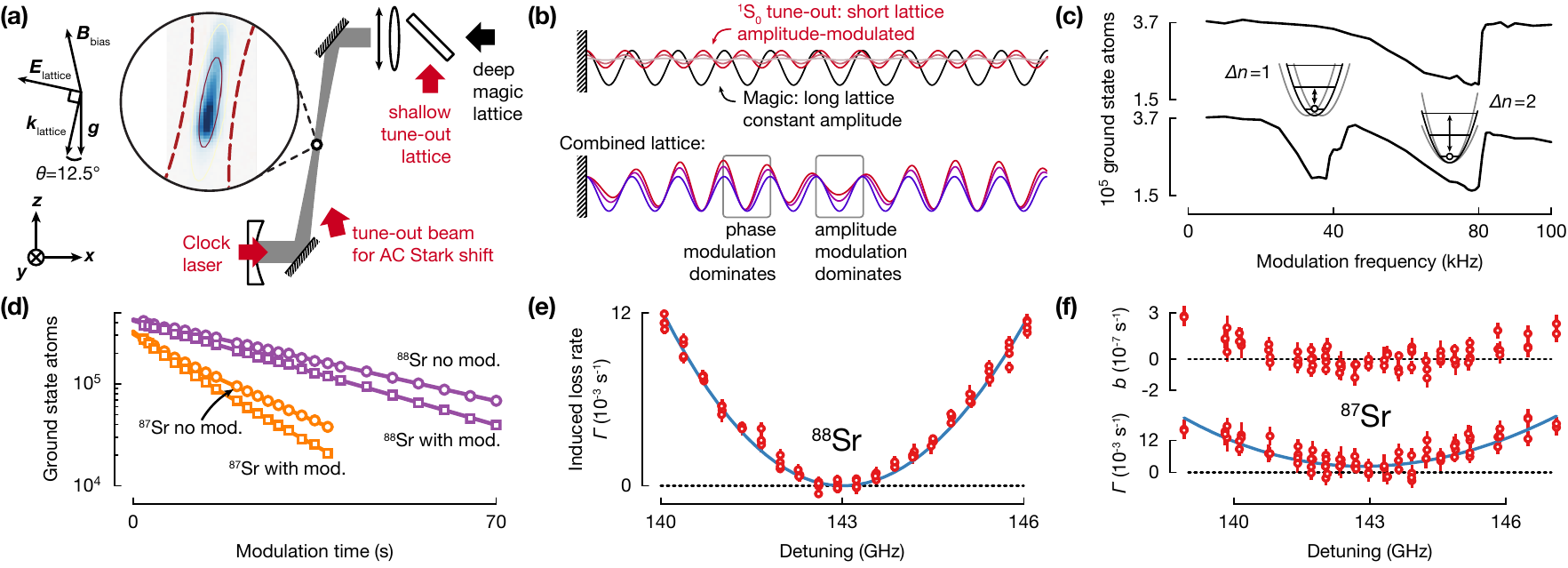}
	\caption{(color online). (a) Experimental setup and optical lattice geometry. We overlap the shallow lattice with the deep magic-wavelength lattice, and trap atoms in this combined trap. (b) When modulating the amplitude of the shallow lattice, the atoms are heated due to phase- and amplitude modulation of the wells of the optical lattice. (c) Lattice modulation spectrum when modulating the magic-wavelength lattice without the shallow lattice (top), and when modulating the shallow lattice (bottom) in the combined trap. (d) Lifetime measurements for bosonic \Sr{88} and fermionic \Sr{87}, when the shallow lattice is modulated or not. (e) The induced exponential loss rate $\Gamma$ has a minimum at the detuning from the \SSZ{}-\TPO{} transition for \Sr{88} corresponding to the tune-out wavelength. (f) For \Sr{87}, interactions lead to an induced two-body loss rate $b$ and an increased uncertainty.}
	\label{fig:tuneout}
\end{figure*}

\paragraph{Measuring the tune-out wavelength.}
We measure the tune-out wavelength by trapping $g$ atoms in a magic-wavelength optical lattice, overlapping an additional optical lattice beam close to the tune-out wavelength and modulating its intensity.
The modulation causes parametric heating and trap loss.
Minimizing the induced loss allows us to precisely determine the tune-out wavelength.
Most measurements of tune-out wavelengths use atom interferometers based on thermal atoms~\cite{lamporesi10,holmgren12,trubko17,copenhaver19} or quantum-degenerate gases~\cite{herold12,leonard15,schmidt16,kao17}.
In contrast, our method is similar to Ref.~\cite{henson15} in that we measure trap loss with an AC method, but with long integration times, reduced systematic effects, and applicability to atoms in excited states, molecules, and trapped ions.

We begin by loading $2 \times 10^5$ $g$ atoms~\cite{snigirev19} into a deep near-vertical one-dimensional optical lattice, created by retro-reflecting \unit{290}{mW} of magic-wavelength laser light at \unit{813.4274(2)}{nm}, as sketched in Fig.~\ref{fig:tuneout}(a).
The magic-wavelength lattice has a longitudinal (transverse) trap frequency of $\sim$\unit{40}{kHz} ($\sim$\unit{100}{Hz}), corresponding to a lattice depth of $k_B \times \unit{5.6}{\mu K} = h\times\unit{116}{kHz}$ and a $1/e^2$ beam waist of \unit{75}{\mu m}, where $k_B$ ($h = 2\pi\hbar$) is the Boltzmann (Planck) constant.
The atoms occupy an ellipsoid with a $1/e^2$ diameter of $\sim$\unit{240}{\mu m} ($\sim$\unit{80}{\mu m}) along the longitudinal (transverse) trap axis, with a typical lattice site filling $\sim$500, as shown in Fig.~\ref{fig:tuneout}(a).
From such \emph{in-situ} and time-of-flight absorption images, we determine atom numbers and temperatures~\cite{snigirev19}.

We then overlap the deep magic-wavelength lattice with a shallow one-dimensional lattice created by retroreflecting \unit{4.5}{mW} of light near the tune-out wavelength, as shown in Fig.~\ref{fig:tuneout}(a).
This geometry allows us to heat the atoms in the combined lattice by intensity-modulating the shallow lattice, as sketched in Fig.~\ref{fig:tuneout}(b).
Since the two lattices are incommensurate there will be lattice sites in which heating due to phase modulation dominates, while in others heating due to amplitude modulation dominates.
By changing the modulation frequency and observing atom loss from the trap~\cite{savard97}, we obtain spectra as shown in Fig.~\ref{fig:tuneout}(c).
We take a reference spectrum (top panel) by modulating the deep lattice intensity, while the shallow lattice is turned off.
In this case, we observe a single minimum in the spectrum at $f_\text{mod} \simeq \unit{80}{kHz}$, corresponding to amplitude modulation and parametric heating~\cite{savard97} that results in transitions between lattice bands that are two motional quanta apart.
The response of the combined lattice due to intensity-modulation of the shallow lattice (bottom panel) shows another minimum at $\sim$\unit{40}{kHz}, corresponding to phase modulation and transitions between adjacent lattice bands~\cite{savard97}.

To compare the effect of heating at different wavelengths of the shallow lattice, we intensity-modulate it at $f_\text{mod}$ and measure the resulting exponential trap loss rate.
Other loss mechanisms such as losses due to intensity noise of both lattices, collisions with background gas atoms, and photon scattering losses, also contribute to the measured heating rate.
We determine the induced trap loss rate $\Gamma(\omega)$ by taking the difference between the measured loss rate without modulation and with modulation.
Examples of such measurements are shown in Fig.~\ref{fig:tuneout}(d) for \Sr{88} (top) and \Sr{87} (bottom).
The \Sr{88} data is well described by an exponential decay because of the isotope's vanishingly small scattering length. In contrast, the \Sr{87} data shows additional superexponential two-body decay.
This decay is due to elastic interactions~\cite{escobar08,stein10} that lead to evaporative trap loss, which we fit with a two-body decay term~\cite{supplemental}.
The induced trap loss rate vanishes when the ground state polarizability $\alpha_g$ crosses zero at the tune-out wavelength, and it is proportional to~\cite{supplemental}
\begin{equation}
\label{eq:scaling}
\Gamma(\omega) \propto \alpha_g^2(\omega) I_{\text{mod}}^2  f_\text{mod}^{-2},
\end{equation}
where $\omega$ is the optical frequency of the tune-out laser, and $I_{\text{mod}}$ is the intensity modulation depth.
To compensate for drifts in $I_\text{mod}$ and the trap frequency, we normalize the measured $\Gamma(\omega)$ according to Eq.~\eqref{eq:scaling}.
The wavelength of the shallow lattice laser is locked to a wavemeter but measured with a self-referenced femtosecond frequency comb, resulting in an absolute frequency error of \unit{3}{MHz}.

The normalized data for \Sr{88} and \Sr{87} are shown in Fig.~\ref{fig:tuneout}(e) and (f), respectively, as a function of detuning from each isotope's \SSZ{}-\TPO{} transition~\cite{miyake19}.
The induced loss rate $\Gamma$ for both isotopes shows a minimum at detuning $\Delta_t$, corresponding to the tune-out wavelength for each isotope.
For \Sr{87}, the induced two-body loss coefficient $b$, given by the difference of the two-body coefficients extracted from the underlying atomic decay curves, shows the same behavior with respect to detuning as $\Gamma$.
This behavior can be explained by an increased tunneling rate in the second lattice band, leading to increased evaporation,
correlated exponential and two-body decay rates, and an increased uncertainty for $\Delta^{87}_t$.
We model the induced loss rate as $\Gamma(\Delta) = c_0 (1-\Delta_{t}/\Delta)^2$~\cite{supplemental}, where $\Delta \equiv \omega - \omega_{\TPO}$ is the detuning from the isotope-shifted \SSZ{}-\TPO{} transition, and the unused fit parameter $c_0$ relates the parametric heating rate to the trap loss rate~\cite{savard97}.
We find $\Delta^{88}_t = 2\pi \times \unit{143.009(8)}{GHz}$ and $\Delta^{87}_t = 2\pi \times \unit{142.86(8)}{GHz}$ for \Sr{88} and \Sr{87}, respectively.
These numbers are in good agreement, considering the empirical two-body loss model for \Sr{87}.
In the Supplemental Material~\cite{supplemental}, we derive a conservative upper limit $|\Delta^{88}_t - \Delta^{87}_t| < 2\pi\times\unit{23}{MHz}$ due to hyperfine splitting, vector, and tensor shifts.
In the following, we use the measured $\Delta^{88}_t$ for \Sr{87} and suppress the superscript for clarity.
This choice leads to a residual $\alpha_g = \pm\unit{0.05}{a.u.}$ from the \unit{2.4}{a.u./GHz} polarizability slope around $\Delta_t$.
Here $\unit{1}{a.u.} = 4\pi\epsilon_0 a_0^3$ is the atomic unit of polarizability, and $\epsilon_0$ ($a_0$) is the vacuum permittivity (Bohr radius).

To minimize systematic shifts in $\Delta_t$ due to laser noise, unsuppressed longitudinal laser modes, and amplified spontaneous emission, we Fourier-filter~\cite{supplemental} the shallow lattice laser and suppress light at the \SSZ{}-\TPO{} transition by >\unit{90}{dB} compared to the carrier.
To avoid saturation of $\Gamma$, and to work with the same $I_\text{mod}$ throughout, we limit the measurement range to a few GHz.
Reducing the measurement range further does not change $\Delta_t$ significantly, and we estimate saturation effects to be negligible.

\begin{figure}
	\centering
	\includegraphics{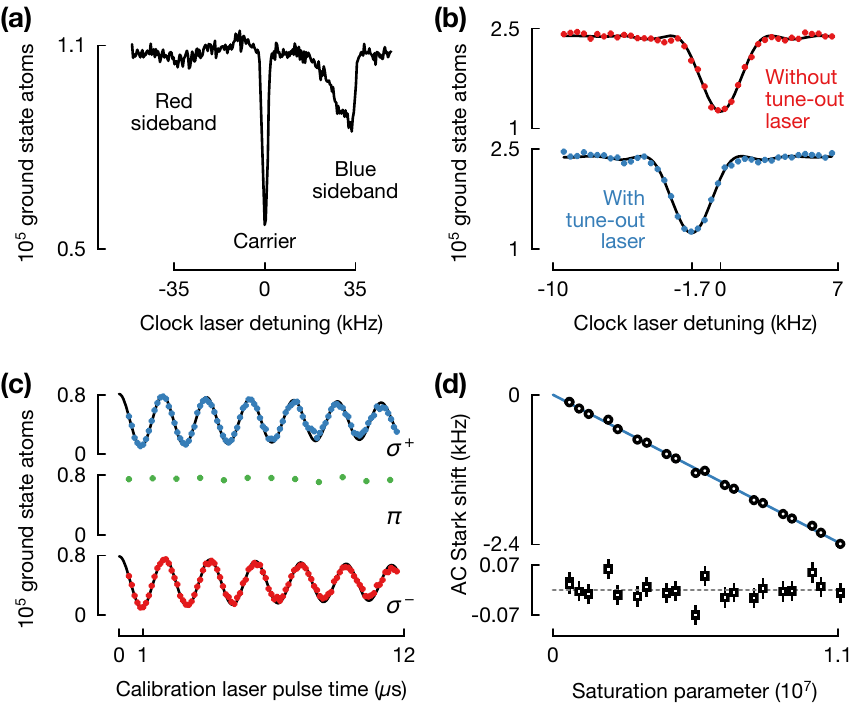}
	\caption{(color online).
		(a) Sideband spectrum of the clock transition in the magic wavelength lattice.
		(b) Typical AC Stark shift spectra obtained on the carrier transition.
		(c) Resonant free-space Rabi oscillations for \Sr{88} calibrate the polarization and intensity of the Stark-shifting beam.
		(d) The Rabi frequencies calibrate the saturation parameter for the measured AC Stark shift (circles). We extract the $e$ state polarizability from the slope of the $\chi_\mathrm{red}^2 = 0.94$ linear fit (solid line). Fit residuals are shown as squares.}
	\label{fig:acstark}
\end{figure}

For \Sr{87}, we observe and fit a statistically significant offset of $2.3(4)\times 10^{-3}~\mathrm{s^{-1}}$ in the induced trap loss rate.
While the offset for \Sr{87} could be explained by contributions from vector and tensor polarizabilities, these cannot occur in \Sr{88}.
When fitting the \Sr{88} data with an offset, we find a much smaller value of $2(1)\times10^{-4}~\mathrm{s^{-1}}$, which also causes a systematic shift of the tune-out frequency of \unit{2}{MHz}.
In conclusion, we find that the tune-out frequency $\omega_t$ for the ground state is detuned from the \SSZ{}-\TPO{} transition at $\omega_\TPO{}$ by
\begin{equation}
  \label{eq:1}
  \Delta_t/2\pi = (\unit{143.009}{GHz}) \pm (\unit{8}{MHz})_\text{stat} \pm (\unit{2}{MHz})_\text{sys}.
\end{equation}

\paragraph{Measuring the excited state polarizability.}
Since the polarizability of the $g$ state vanishes at $\omega_t$, the AC Stark shift on the clock transition induced by a laser beam at $\omega_t$ is solely determined by the excited state polarizability.
For this reason, we can directly measure the $e$ state polarizability $\alpha_e(\omega_t)$ by Stark shift spectroscopy.
For these measurements, we prepare a sample of \Sr{87}
in a one-dimensional magic wavelength lattice.
We propagate a clock laser beam at 3.5$^\circ$ with respect to the lattice axis and overlap it with the lattice at the position of the atoms.
A typical spectrum of the clock transition is shown in Fig.~\ref{fig:acstark}(a), consistent with a $\sim$\unit{1}{\mu K} temperature~\cite{blatt09}, confirmed by time-of-flight data.
On the carrier transition, we observe damped Rabi oscillations, consistent with the mismatch between the lattice axis and the clock laser wave vector~\cite{blatt09}.
For the following measurements, we illuminate the atoms with the clock laser for \unit{0.3}{ms}, corresponding to the maximum excited state fraction.
For the AC Stark shift spectroscopy, we additionally apply a laser beam at $\omega_t$ with a $1/e^2$ waist of \unit{300}{\mu m}, as illustrated in Fig.~\ref{fig:tuneout}(a).
As a function of clock laser detuning, we observe the spectra  shown in Fig.~\ref{fig:acstark}(b), and fit them to extract the center frequencies.
To calibrate the intensity, we load a sample of \Sr{88} into the same magic-wavelength lattice at the same position as the \Sr{87} sample.
After diabatically switching off the lattice, we measure free-space Rabi oscillations on each of the three \SSZ{}-\TPO{} transitions in a small magnetic bias field, as shown in Fig.~\ref{fig:acstark}(c).
We fit the Rabi oscillations with an analytic solution to the optical Bloch equations~\cite{loudon00}, and extract the Rabi frequency $\Omega_\pm$ for each $\sigma^\pm$ polarization component, for an applied power $P_0=\unit{112}{\mu W}$.
The Rabi frequency of the Stark-shifting beam with power $P$  is calibrated as $\Omega^2 \equiv (\Omega_+^2 + \Omega_-^2) P / P_0$, which allows expressing the AC Stark shift of the clock transition at the tune-out frequency $\Delta \omega_{eg} = - \omega^3_\TPO{} \tau_\TPO{} \alpha_e(\omega_t) \Omega^2 / (12 \pi \epsilon_0 c^3)$ in terms of measured quantities, where $c$ is the speed of light.
In Figure \ref{fig:acstark}(d), we plot $\Delta\omega_{eg}$ as a function of the saturation parameter $s_0 = 2\Omega^2\tau_{\TPO}^2$, and use a linear fit to extract the excited state polarizability
\begin{equation}
  \label{eq:alphae-result}
  \alpha_e(\omega_t) = (1555 \pm 8_\mathrm{stat} \pm 2_\mathrm{sys})~\mathrm{a.u.}
\end{equation}
of \Sr{87} at the tune-out frequency, where the systematic uncertainty includes the mismatch between the tune-out frequency for \Sr{88} and \Sr{87}, but is dominated by the effect of the laser spectrum.
Our measured polarizability agrees well with our theoretical prediction of \unit{1546(14)}{a.u.}, based on Ref.~\cite{safronova13}.

\begin{figure}
  \centering
  \includegraphics{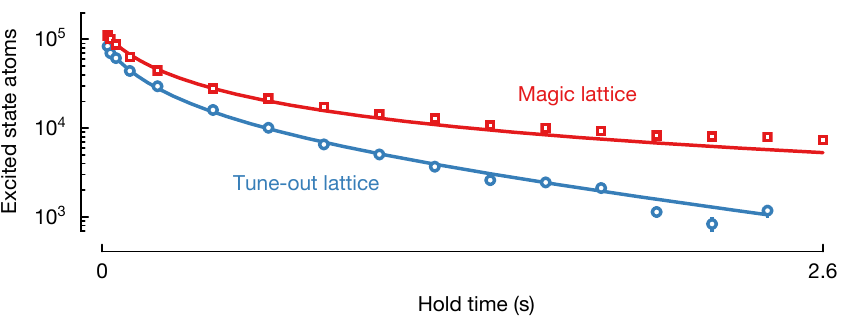}
  \caption{(color online). Number of $e$ atoms versus hold time in a one-dimensional optical lattice at the magic wavelength (squares) and at the tune-out wavelength (circles). Both lattices exhibit the expected losses due to inelastic collisions, while the tune-out lattice additionally exhibits exponential one-body decay due to light scattering.}
  \label{fig:lattice}
\end{figure}

\paragraph{Trapping excited atoms at the tune-out wavelength.}
We prepare a sample of $e$ state \Sr{87} atoms in the magic-wavelength lattice and transfer them to the tune-out lattice~\cite{supplemental}.
In Fig.~\ref{fig:lattice}, we compare the number of $e$ atoms in the tune-out lattice as a function of hold time to the case where we trap $e$ atoms in the magic-wavelength lattice.
In both cases, the atoms decay superexponentially via $e$-$e$ collisions~\cite{bishof11}.
In the magic wavelength lattice, this inelastic loss dominates, while the atoms in the tune-out lattice experience additional exponential loss with a $1/e$ lifetime $\sim$\unit{1.2}{s}.
The intensity of the tune-out lattice is chosen to match the trap frequencies of the magic-wavelength and tune-out lattices at $\sim$\unit{40}{kHz}, corresponding to a lattice depth of $\sim$$17\times \hbar\omega_\mathrm{rec}$ ($\omega_\mathrm{rec}/2\pi = \unit{4.8}{kHz}$ is the lattice recoil frequency), confirmed by parametric heating.
The measured lifetime agrees well with the theoretically predicted loss due to photon scattering for each lattice axis of \unit{24}{s} per recoil of lattice depth.
Depending on the application, a compromise between lattice depth and tunneling rate needs to be found.
For instance, a two-dimensional tune-out lattice trapping $e$ atoms in a Mott insulator state would have a lifetime $\sim$\unit{1}{s}.

\paragraph{Determination of atomic lifetimes.}

The $g$ state polarizability $\alpha_g(\omega) = \alpha_g(\SPO;\omega) + \alpha_g(\TPO;\omega) + \alpha_\text{vc}(\omega)$ is dominated by coupling to the $\SPO{}$ and $\TPO{}$ states.
All other contributions from valence and core electronic states are small and we calculate their combined value $\alpha_\text{vc}$ with percent-level precision~\cite{safronova13}.
Since both \SPO{} and \TPO{} decay only to $g$, and because the tune-out frequency is far-detuned with respect to each state's natural linewidth, each contribution $\alpha_g(j;\omega)\propto\tau_j^{-1}$~\cite{supplemental}, where $\tau_j$ is the state's natural lifetime.
At the tune-out frequency $\alpha_g$ vanishes, leading to a strong constraint on the relationship between the lifetimes $\tau_\SPO{}$ and $\tau_\TPO{}$.
While $\tau_\TPO{}$ has been recently measured directly~\cite{nicholson15}, the currently accepted value for $\tau_\SPO{}$ comes from photoassociative spectroscopy~\cite{nagel05,yasuda06}.
Using our tune-out wavelength and
$\tau_\TPO{}$~\cite{nicholson15}, we find $\tau_\SPO{}$ = \unit{5.234(8)}{ns}~\cite{supplemental}, a 7$\sigma$ discrepancy with the currently accepted value~\cite{yasuda06}.
The $e$ state polarizability $\alpha_e$ at the tune-out wavelength is dominated (87\%) by the \TPZ{}-\TSO{} transition.
We determined all other contributions with a total uncertainty of \unit{4}{a.u.} using a high-precision relativistic method~\cite{safronova13}.
Combining these theoretical values with our measurement of $\alpha_e$ determines the \TPZ{}-\TSO{} matrix element.
The \TSO{} state dominantly decays to the \TPJ{} levels, while its decay to \SPO{} is negligible at the present level of accuracy.
We calculated the \TSO{}-\TPJ{} branching ratios with 0.1\% accuracy, which allows us to extract $\tau_{\TSO{}}$ from the \TPZ{}-\TSO{} matrix element.
This accurate prediction of branching ratios is possible due to very similar electronic correlation effects for these transitions, which largely cancel for their ratios.
We find $\tau_\TSO{} = \unit{13.92(11)}{ns}$, an improvement of an order of magnitude over prior measurements that ranged from \unit{10.9(1.1)}{ns} to \unit{15.0(8)}{ns}~\cite{brinkmann69,havey77,jonsson84}.
We note in passing that the currently best values for $\tau_\TPO{}$ and the lifetime of the \TDO{} state are correlated because they are extracted from a single data set~\cite{nicholson15}.
The $\TDO{}$ lifetime directly determines the dynamic contribution to the Sr lattice clock blackbody radiation shift~\cite{safronova13}, its currently largest systematic uncertainty~\cite{bothwell19}.
This uncertainty can be directly improved by a new direct measurement of $\tau_\SPO{}$ in combination with our results and Ref.~\cite{nicholson15}.
Our measurements show that direct measurements of atomic lifetimes and improvements to atomic structure calculations will be essential in bringing optical frequency standards to the $10^{-19}$ level.

In conclusion, we have demonstrated state-dependent optical lattices for the clock states of strontium at the tune-out wavelength for its ground state.
With a new spectroscopic method, we achieved a record suppression of the lattice depth for the ground state of more than four orders of magnitude.
Our method can be applied to thermal gases of atoms, molecules~\cite{bause19}, or to trapped ions.
As a modulation technique for trapped particles, the method benefits from suppression of systematic errors and long integration times.
Using our technique in three-dimensional optical lattices in combination with band mapping~\cite{blatt15} will enable measurements of excited state tune-out wavelengths, such as the tune-out wavelength for the \TPZ{} state around \unit{633}{nm}~\cite{safronova15}, even in the presence of interactions.
We have demonstrated high-fidelity, state-dependent control of the strontium optical qubit.
Combining our results with single-site addressing and control~\cite{weitenberg11} removes the main obstacle for the realization of quantum computation and quantum simulation schemes with two-electron atoms~\cite{daley11b}.
Finally, our work creates new opportunities to use state-dependent optical lattices for quantum simulations of nanophotonics~\cite{devega08,krinner18,tudela18} and quantum
chemistry~\cite{arguello19}.

\begin{acknowledgments}
  We thank J.~Ye, A.~Gonz\'alez-Tudela, J.~I.~Cirac, and M.~J.~Martin for stimulating discussions.
  This work was supported by funding from the European Union (UQUAM Grant No. 319278 and PASQuanS Grant No. 817482).
  A.\,J.\,P. was supported by a fellowship from the Natural Sciences and Engineering Research
Council of Canada (NSERC), funding ref. no. 517029, and N.\,\v{S}. was supported by a Marie Sk\l{}odowska-Curie individual fellowship, grant agreement no. 844161.
Work at U. Delaware was performed under the sponsorship of the U.S. Department of Commerce, National Institute of Standards and Technology.
S.\,G.\,P. acknowledges support by the Russian Science Foundation under Grant No.~19-12-00157.
\end{acknowledgments}


%

\end{document}


\title{Supplemental Material: \\
  State-dependent optical lattices for the strontium optical qubit}

\author{A. Heinz}
\thanks{A.\,H. and A.\,J.\,P. contributed equally to this work.}
\author{A. J. Park}
\thanks{A.\,H. and A.\,J.\,P. contributed equally to this work.}
\author{N. \v{S}anti\'c}
\author{J. Trautmann}
\affiliation{
  Max-Planck-Institut f{\"u}r Quantenoptik,
  Hans-Kopfermann-Stra{\ss}e 1,
  85748 Garching, Germany}
\affiliation{
  Munich Center for Quantum Science and Technology,
  80799 M{\"u}nchen, Germany}
\author{S. G. Porsev}
\affiliation{Department of Physics and Astronomy, University of Delaware, Newark, Delaware 19716, USA}
\affiliation{Petersburg Nuclear Physics Institute of NRC ``Kurchatov Institute,'' Gatchina, Leningrad district 188300, Russia}
\author{M. S. Safronova}
\affiliation{Department of Physics and Astronomy, University of Delaware, Newark, Delaware 19716, USA}
\affiliation{Joint Quantum Institute, National Institute of Standards and Technology and the University of Maryland, College Park, Maryland, 20742, USA}
\author{I. Bloch}
\affiliation{
  Max-Planck-Institut f{\"u}r Quantenoptik,
  Hans-Kopfermann-Stra{\ss}e 1,
  85748 Garching, Germany}
\affiliation{
  Munich Center for Quantum Science and Technology,
  80799 M{\"u}nchen, Germany}
\affiliation{
  Fakult{\"a}t f{\"u}r Physik,
  Ludwig-Maximilians-Universit{\"a}t M{\"u}nchen,
  80799 M{\"u}nchen, Germany}
\author{S. Blatt}
\email{sebastian.blatt@mpq.mpg.de}
\affiliation{
  Max-Planck-Institut f{\"u}r Quantenoptik,
  Hans-Kopfermann-Stra{\ss}e 1,
  85748 Garching, Germany}
\affiliation{
  Munich Center for Quantum Science and Technology,
  80799 M{\"u}nchen, Germany}

\date{\today}

\maketitle

\section{Parametric heating in incommensurate lattices}
\label{sec:incommensurate_lattices}

The optical dipole potential of two retroreflected laser beams with different wavelengths $\lambda_i = 2\pi/k_i$ along the $z$-direction is
\begin{equation}
  \label{eq:appB1}
	u(z) = u_{1}\text{cos}^{2}(k_{1}z)+u_{2}(t)\text{cos}^{2}(k_{2}z).
  \end{equation}
Here $u_i$ is the lattice depth induced on the atomic state via the AC Stark effect.
In the parametric-heating experiments described in the main text, the first lattice is the deep magic-wavelength lattice with a fixed depth $u_1$.
The second lattice is the shallow tune-out lattice with a variable depth $u_2(t) = u_2^0 + u_2^\mathrm{mod}\cos(\omega_\mathrm{mod} t)$ that is sinusoidally modulated around its mean value $u_2^0$ with a modulation amplitude $u_2^\mathrm{mod}$ and a modulation frequency $\omega_\mathrm{mod}$.
Since we use shallow tune-out lattices with $u_2(t) \ll u_1$, we treat the effect of the shallow lattice in perturbation theory by expressing its effect as a variation in the trap position and the trap frequency of each lattice site~\cite{savard97,gehm98,gehm00}.
To this end, we expand the combined lattice potential around the position $z_j = (2j+1)\pi/(2k_1)$ of the $j$-th minimum of the deep  lattice and find
\begin{equation}
  \label{eq:appB2}
  \begin{aligned}
    u(z) & \approx u_2(t) \cos^2(k_2 z_j)
    - u_2(t) \sin(2k_2 z_j) k_2 (z-z_j) \\
    & \qquad + [u_1 k^2_1 - u_2(t) k^2_2 \cos(2k_2 z_j)]
    (z-z_j)^{2}. \\
  \end{aligned}
\end{equation}
By completing the squares and dropping constant terms, we can bring the above expression into the standard form~\cite{savard97,gehm98,gehm00} for parametric heating in a harmonic oscillator
\begin{equation}
  \label{eq:2}
  u \approx \frac{m\omega_j^2}{2}(1 + \epsilon)[z - (\tilde{z}_j + \delta_j)]^2,
\end{equation}
where $m$ is the mass of the atom, $\omega_j$ ($\tilde{z}_j$) is the unmodulated trap frequency (position of the minimum) of lattice site $j$, and $\epsilon$ ($\delta_j$) is the amplitude (position) modulation causing transitions between the harmonic oscillator levels.
Explicitly, we find
\begin{equation}
  \label{eq:3}
  \begin{aligned}
    m\omega_j^2/2 &= u_1 k_1^2 - u_2^0 k_2^2 \cos 2k_2 z_j, \\
    \epsilon &= \frac{u_2^\mathrm{mod}}{u_1} \cos\omega_\mathrm{mod} t, \\
    \tilde{z}_j &= z_j - \left(\frac{k_2}{2k_1^2}\sin 2k_2 z_j\right) \frac{u_2^0}{u_1}, \\
    \delta_j &= \left(\frac{k_2}{2k_1^2}\sin 2k_2 z_j\right) \frac{u_2^\mathrm{mod}}{u_1} \cos\omega_\mathrm{mod} t. \\
  \end{aligned}
\end{equation}
From these expressions we see that the trap frequency of the deep lattice $m \omega_1^2/2 = u_1 k_1^2$ is very weakly influenced by the presence of a weak lattice with bare trap frequency $m \omega_2^2/2 = u_2^0 k_2^2$.
These trap frequencies scale with the ratio of the ground state polarazibilites $\alpha_g(\lambda_i)$ at the magic and tune-out wavelengths and the lattice powers $P_i$.
For our parameters, we find
\begin{equation}
  \label{eq:5}
  \frac{\omega_2^2}{\omega_1^2} \propto \frac{\alpha_g(\lambda_2) P_2}{\alpha_g(\lambda_1) P_1}
  \simeq 5\times 10^{-5}.
\end{equation}
For this reason, the trap frequency of the combined lattice is unchanged when modulating the deep lattice by itself or when modulating the shallow lattice in the presence of the deep lattice.
In the latter case, the modulation spectrum simply acquires a second minimum at the trap frequency, corresponding to position modulation of the lattice sites.

\section{Trap loss fitting}
\label{sec:trap-loss}

To analyze the trap loss curves throughout the main text, we fit the atom number data $N(t)$ as a function of time $t$ with exponential decay curves $N(t) = N_0 e^{-\Gamma t}$ resulting from a single-body decay process $\dot{N} = -\Gamma N$.
We decide whether to fit with superexponential decay
\begin{equation}
  \label{eq:7}
  N(t) = N_0 \frac{\Gamma e^{-\Gamma t}}{\Gamma + N_0 b (1 - e^{-\Gamma t})},
\end{equation}
due to an additional two-body decay process $\dot{N} = -\Gamma N - b N^2$ based on a $\chi^2$ test.
This two-body loss model assumes a homogeneous density in the trap that attenuates homogeneously without changing the sample temperature.
Because we do not take the temperature effects of evaporation and anti-evaporation into account, the results of the superexponential decay curves are \emph{not} used to derive any of the important quantities in the main text and are meant simply as a guide to the eye.

\section{Trap loss rate rescaling}
\label{sec:rescaling}

If we modulate with $\omega_\mathrm{mod} = 2\pi f_\mathrm{mod} = 2\omega_1$ (at twice the deep lattice trap frequency), the differential trap loss rate is proportional to the heating rate caused by the intensity modulation alone~\cite{savard97,gehm98,gehm00}
\begin{equation}
	\label{eq:appC1}
	\begin{aligned}
	\Gamma_\text{mod}&\propto f^{2}_{\text{mod}}S_{\epsilon}(f_{\text{mod}})\\
	&\propto\omega^{2}_1 \left(\frac{u_2^\mathrm{mod}}{u_1}\right)^2 \\
	\end{aligned}
\end{equation}
where $S_{\epsilon}(f_\text{mod})$ is the power spectrum of the fractional intensity noise at the modulation frequency $f_\text{mod} = \omega_\mathrm{mod}/2\pi$.
The ratio of shallow modulation amplitude to deep lattice trap depth, $u_2^\mathrm{mod}/u_1 = \alpha_g(\lambda_2) I_\mathrm{mod}/[\alpha_g(\lambda_1) I_1]$ where $I_\mathrm{mod}$ is the intensity modulation amplitude of the shallow lattice and $I_{1}$ is the intensity of the deep lattice.
Thus,
\begin{equation}
	\label{eq:appC2}
	\Gamma_\text{mod} \propto \alpha_g(\lambda_2)^2 f^2_\mathrm{mod} \left(\frac{I_\mathrm{mod}}{I_1}\right)^2
	\propto \alpha_g(\lambda_2)^2 \frac{I^2_\mathrm{mod}}{f^2_\mathrm{mod}},
\end{equation}
where we have used $I_{1} \propto \omega_1^2 \propto f^2_\mathrm{mod}$.
To compensate for possible experimental variations in $\omega_1$ and $I_\mathrm{mod}$, we scale the measured induced loss rate with respect to the reference values
\begin{equation}
	\label{eq:appC3}
	\begin{aligned}	\Gamma^{\text{scaled}}_{\text{mod}}&=\Gamma_{\text{mod}}\bigg(\frac{f_{\text{mod}}}{f^{\text{ref}}_{\text{mod}}}\bigg)^2\bigg(\frac{I^{\text{ref}}_{\text{mod}}}{I_{\text{mod}}}\bigg)^2,\\
	\end{aligned}
\end{equation}
and work with these scaled induced loss rates throughout the main text.

\section{The \SSZ{} ground state tune-out wavelength}
\label{sec:theory}

The polarizability of an electronic state consists of contributions from the core electrons and valence electrons.
The valence part of the ground state \SSZ{} ($g$) polarizability is determined by summing over all the contributions of excited states dipole-coupled to $g$, dominated by the \SPO{} and \TPO{} contributions.
For this reason, we write the $g$ polarizability using the four components
\begin{equation}
  \label{eq:appA1}
  \alpha_g(\omega) = \alpha_g(\SPO;\omega) + \alpha_g(\TPO;\omega) + \alpha_g(v; \omega) + \alpha_g(c; \omega),
\end{equation}
where $\alpha_g(j;\omega)$ are the contributions from the excited state $j$, $\alpha_g(v; \omega)$ is the sum of contributions from all other valence states, $\alpha_g(c; \omega)$ is the contribution of the core electrons, and $\omega = 2\pi c/\lambda$ is the optical frequency for wavelength $\lambda$.

The contribution from the \SPO{} and \TPO{} states further splits into scalar, vector, and tensor components as~\cite{manakov86,lekien13}
\begin{equation}
  \label{eq:appA2}
  \begin{aligned}
  \alpha_g(j;\omega) & = \alpha_g^s(j;\omega) \\
  & +  \alpha_g^v(j;\omega) (i\vec{\epsilon}\times\vec{\epsilon}^*)\cdot\vec{e}_z  \frac{m_{F}}{F} \\
  & + \alpha_g^t(j;\omega) \frac{3|\vec{\epsilon}\cdot\vec{e}_z|^2 - 1}{2}
  \frac{3m^2_{F} -F(F+1)}{F(2F-1)}, \\
  \end{aligned}
\end{equation}
where $F$ is the hyperfine quantum number associated with the  ground state $g$, $m_F$ is the corresponding magnetic quantum number, and $j$ indicates one of the excited states \SPO{} and \TPO{}.
The \SSZ{} ($g$) state of the fermionic isotope \Sr{87} has a single hyperfine state $F=9/2$, since the electronic angular momentum $J$ and the nuclear spin $I$ are $0$ and $9/2$, respectively ($|J-I|\leq F \leq |J+I|$).
For the bosonic isotope \Sr{88} where $I=0$, $J$ replaces $F$ ($F=J$).
Therefore, any discussion involving $F_{j}$ throughout this text only applies to \Sr{87}.
The quantization axis is assumed to be along $\vec{e}_z$, and $\vec{\epsilon}$ is the complex polarization vector of the applied laser at frequency $\omega$.
The vector and tensor components depend on the polarization of the applied beam, while the scalar component does not.

The scalar, vector, and tensor parts themselves can be written as sums over all hyperfine states $F_j$ in the excited fine structure state~\cite{manakov86,lekien13}
\begin{equation}
  \label{eq:appA3}
  \begin{aligned}
    \alpha_g^s(j;\omega) & = \sum_{F_j} \frac{2}{3} g^{(0)}_{j, F_j}(\omega) |D_{j,F_j}|^2,\\
    \alpha_g^v(j;\omega) & = \sum_{F_j} (-1)^{F+F_j+1}\sqrt{\frac{6F(2F+1)}{F+1}}\\
    & \qquad\qquad\times \begin{Bmatrix}
      1 & 1 & 1 \\ F & F & F_j \\
    \end{Bmatrix}
     g^{(1)}_{j,F_j}(\omega) |D_{j,F_j}|^2, \\
    \alpha_g^t(j;\omega) & = \sum_{F_j} (-1)^{F+F_j} \sqrt{\frac{40F(2F-1)(2F+1)}{3(F+1)(2F+3)}} \\
        & \qquad\qquad \times \begin{Bmatrix}
      1 & 1 & 2 \\ F & F & F_j \\
    \end{Bmatrix}
    g^{(2)}_{j,F_j}(\omega) |D_{j,F_j}|^2. \\
  \end{aligned}
\end{equation}
Scaling of the polarizability terms as a function of laser detuning is encapsulated in the detuning factors
\begin{equation}
  \label{eq:appA4}
  \begin{aligned}
  g^{(K)}_{j,F_j}(\omega) &= \frac{1}{2\hbar}\text{Re}\bigg(\frac{1}{\omega_{j,F_j}-\omega-\frac{i}{2\tau_{j}}}+\frac{(-1)^{K}}{\omega_{j,F_j}+\omega+\frac{i}{2\tau_{j}}}\bigg)\\
  &\approx \frac{1}{2\hbar}\bigg(\frac{1}{\omega_{j,F_j}-\omega}+\frac{(-1)^{K}}{\omega_{j,F_j}+\omega}\bigg)\\
  \end{aligned}
\end{equation}
where $\omega_{j,F_j}$ is the transition frequency from the ground to the excited hyperfine state $F_j$, $\tau_{j}$ is the excited state lifetime, and $K=0$, $1$, $2$ for scalar, vector, and tensor polarizabilities, respectively~\cite{manakov86,lekien13}.
Because we work in the far-detuned regime, we can ignore the imaginary terms in the denominators of Eqn.~\eqref{eq:appA4}, which we have confirmed numerically.
The detuning factor is identical for the scalar and tensor polarizabilities, but   the counter-rotating term for the vector polarizability changes its sign.
Note that $\omega_{j,F_j}$ can be expressed in terms of the (hypothetical) hyperfine-free transition frequency $\bar{\omega}_{j}$ and the hyperfine shift $\Delta_{j,F_{j}}$ as $\omega_{j,F_j}=\bar{\omega}_{j}+\Delta_{j,F_{j}}$.
The hyperfine shift can be calculated from the magnetic dipole interaction constant $A_j$ and the electric quadrupole interaction constant $Q_j$ as~\cite{boyd07,boyd07b}
\begin{equation}
  \label{eq:8}
  \begin{aligned}
    \Delta_{j,F_j} &= \frac{A_j}{2}K_j + \frac{Q_j}{4}
    \frac{\frac{3}{2}K_j(K_j+1) -2I(I+1)J_j(J_j+1)}{I(2I-1)J_j(2J_j-1)}, \\
    K_j &= F_j(F_j+1) - J_j(J_j+1) - I(I+1),
  \end{aligned}
\end{equation}
where $J_j$ is the electronic angular momentum of the excited state.
We use the hyperfine constants $A_\SPO{} = -2\pi\times\unit{3.4(4)}{MHz}$, $Q_\SPO{} = 2\pi\times\unit{39(4)}{MHz}$, $A_\TPO{} = -2\pi\times\unit{260.084(2)}{MHz}$, and $Q_\TPO{} = -2\pi\times\unit{35.658(6)}{MHz}$, summarized in Ref.~\cite{boyd07}.
The isotope shift of $\bar{\omega}_\TPO{}$ was measured recently in Ref.~\cite{miyake19}: the hyperfine-free transition frequency of \SSZ{}-\TPO{} in \Sr{87} is red detuned by \unit{62.1865(123)}{MHz} from the transition frequency in \Sr{88}, \unit{434,829,121,311(10)}{kHz}~\cite{ferrari03}.
The \SSZ{}-\SPO{} transition frequency does not need to be as precise and is obtained from Ref.~\cite{sansonetti10}.
The isotope shift on this transition does not influence the calculations, as will be shown below.

All polarizability terms scale with the modulus-squared of the reduced matrix element associated with the  corresponding dipole transition
\begin{equation}
  \label{eq:appA5}
  |D_{j,F_j}|^2 = |\langle g, F\| D \| j, F_j \rangle|^2.
\end{equation}
Since the \SSZ{} state of \Sr{87} has a single hyperfine state, the reduced matrix element can be expressed in terms of the excited state lifetime $\tau_j$ as
\begin{equation}
  \label{eq:appA6}
  |D_{j,F_j}|^2 = \frac{3\pi\epsilon_0\hbar c^3}{\omega^3_{j,F_j}\tau_j} (2 J_j+1)(2F_j+1)
  \begin{Bmatrix}
    J & J_j & 1 \\ F_j & F & I \\
  \end{Bmatrix}^2.
\end{equation}
The two excited states of interest \SPO{} and \TPO{} both have $J_j=1$, and thus have the same three hyperfine states $F_j = 7/2$, $9/2$, and $11/2$. For this reason, the scalar, vector, and tensor polarizabilities only differ in the group-theoretic numerical prefactors associated with the excited states, except for the slightly different detuning dependence of the vector polarizability.

The situation simplifies strongly for the case of \Sr{88}.
If we replace $F$ ($F_j$) with $J$ ($J_j$) in Eqn.~\eqref{eq:appA3}, the vector and tensor polarizabilities vanish.
Only the scalar component remains in the absence of hyperfine structure.

\begin{figure*}
  \centering
  \includegraphics{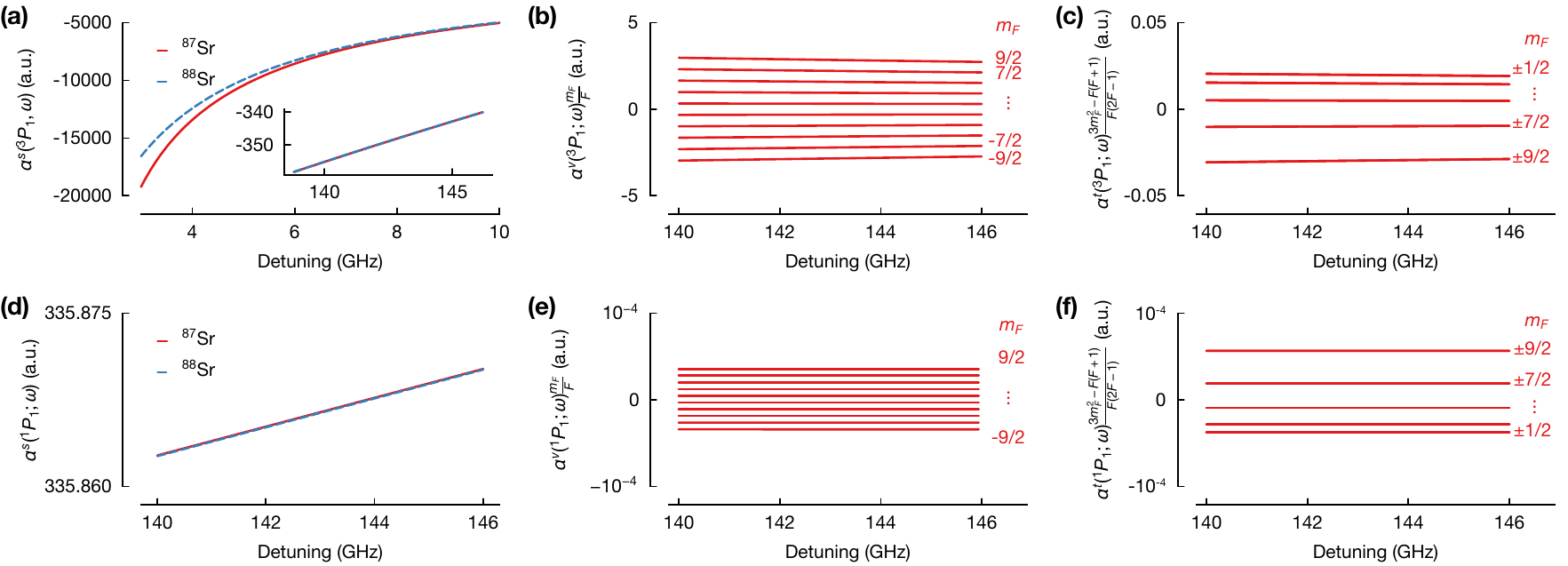}
  \caption{(color online). (a) The \TPO{} scalar contribution as a function of detuning from the \SSZ{}-\TPO{} transition. For the case of \Sr{87}, the detuning is referenced with respect to $\bar{\omega}_{\TPO}$, which includes the isotope shift. The inset shows $\alpha^s_{\TPO}$ at the detuning range relevant to our data. (b) the \TPO{} vector contribution as a function of detuning from the \SSZ{}-\TPO{} transition for different $m_{F}$ states (Eqn.~\eqref{eq:appA3}). We assume the beam ellipticity of 1 to show the upper limit. (c) the \TPO{} tensor contribution (Eqn.~\eqref{eq:appA3}) as a function of detuning from the \SSZ{}-\TPO{} transition for different $m_{F}$ states. Here, we assume a linear polarization to show the upper limit. (c)-(f) same as (a)-(c), but from the \SPO{} contributions.
}
  \label{fig:appendixA}
\end{figure*}

To study the contributions from each component in detail, we dig deeper into the scalar, vector, and tensor polarizabilities of the
\Sr{87} \SSZ{} state. By specializing the prefactors, we re-write the reduced matrix elements as
\begin{equation}
  \label{eq:appA6}
  \begin{aligned}
    |D_{j,F_j}|^2 & = \frac{3\pi\epsilon_0 \hbar c^3}{\omega^3_{j,F_{j}}\tau_j} \{4/5, 1, 6/5\},\\
  \end{aligned}
\end{equation}
where the numerical factors in the braces correspond to $F_j = 7/2$, $9/2$, and $11/2$, respectively.
As expected, this means that the lifetime $\tau$ proportionally scales up scalar, vector, and tensor polarizabilities. Bringing it all together, we find
\begin{equation}
  \label{eq:appA8}
  \begin{aligned}
    \alpha_g^s(j;\omega) & = 2\sum_{F_j} \frac{3\pi\epsilon_0 \hbar c^3}{\omega^3_{j,F_{j}}\tau_j}\left\{\frac{4}{15},\frac{5}{15},\frac{6}{15}\right\} g^{(0)}_{j,F_j}(\omega), \\
    \alpha_g^v(j;\omega) & = \sum_{F_j}\frac{3\pi\epsilon_0 \hbar c^3}{\omega^3_{j,F_{j}}\tau_j}
    \left\{-\frac{44}{55},-\frac{10}{55},\frac{54}{55}\right\} g^{(1)}_{j,F_j}(\omega), \\
    \alpha_g^t(j;\omega) & = \frac{\sqrt{2}}{3}\sum_{F_j} \frac{3\pi\epsilon_0 \hbar c^3}{\omega^3_{j,F_{j}}\tau_j}
    \left\{-\frac{88}{165}, \frac{160}{165},-\frac{72}{165}\right\} g^{(2)}_{j,F_j}(\omega). \\
  \end{aligned}
\end{equation}
To get an intuitive picture of what happens in the far-detuned regime compared to the hyperfine structure, we approximate $|D_{j,F_j}|^2 \approx \frac{3\pi\epsilon_0 \hbar c^3}{\bar{\omega}^3_j\tau_j} \{4/5, 1, 6/5\} $ to pull $\frac{3\pi\epsilon_0 \hbar c^3}{\omega^3_{j,F_{j}}\tau_j}$ out of the sum in Eqn.~\eqref{eq:appA8}.
After this approximation, we see that both vector and tensor polarizabilities sum to zero when we are in the far-detuned regime compared to the hyperfine structure, where the detuning factor contributes equally, and can be pulled out of the sum.
In this regime, the only contribution that survives is the scalar polarizability.
For the same reason, we also expect negligible differences in $\alpha^s$ between the two isotopes in this regime.
The scalar polarizabilities of the two isotopes are illustrated in Fig.~\ref{fig:appendixA}(a) where we see that they become indistinguishable as the detuning from the \SSZ{}-\TPO{} transition increases (the plot was generated based on Eqn.~\eqref{eq:appA8} without the approximation on $|D_{j,F_j}|^2$).
Note that the detuning for \Sr{87} is referenced with respect to $\bar{\omega}_{j}$, which takes the isotope shift into account.

Next, we take a closer look at each contribution in the detuning range relevant to our experimental data.
Comparing Fig.~\ref{fig:appendixA}(b)-(c) and (e)-(f), the \SPO{} tensor and vector contributions are more than two orders of magnitude smaller than the contributions from the \TPO{} state, since the frequency of the shallow lattice is more than a few hundred THz detuned from the \SSZ{}-\SPO{} transition.
For this reason, we can ignore the vector and tensor contributions to $\alpha_g({\SPO};\omega)$ and set $\alpha_g({\SPO};\omega)=\alpha_g^{s}({\SPO};\omega)$.

With this simplification, we analyze the difference of the \Sr{87} and \Sr{88} tune-out detuning $\Delta_t$ due to the hyperfine structure.
At the tune-out detuning, $\alpha_g(\SPO{};\omega)$ balances $\alpha_g(\TPO{};\omega)$ and the remaining valence and core contributions to the polarizability, $\alpha_{\text{vc}} = \alpha_g(v; \omega) + \alpha_g(c; \omega)$.

We calculated the valence part of $\alpha_g$ by solving the inhomogeneous equation as described in Refs.~\cite{kozlov99,safronova13}.
Then, using the sum-over-states formula Eqn.~\eqref{eq:appA3}, we extracted the contributions of the \SPO{} and \TPO{} states and determined the remaining valence contributions, $\alpha_g(v,\Delta_{t})= \unit{6.57(14)}{a.u.}$
The core part of the polarizability, $\alpha_g(c;\Delta_{t})$, was calculated in the single-electron approximation including random-phase approximation corrections~\cite{safronova99} to be $\alpha_g(c;\Delta_{t}) = \unit{5.30(5)}{a.u.}$

Table~\ref{table_appendixA1} shows the shifts due to scalar, vector, and tensor polarizabilities.
For the calculations due to the vector polarizabilities, we used an upper limit on the beam ellipticity of 2\%, measured by a polarimeter after transmission through our vacuum chamber, and assumed perfect alignment of $\vec{\epsilon}\cdot\vec{e}_z = 1$ to evaluate the tensor polarizability.
The shifts depend on the $m_F$ states (Eqn.~\eqref{eq:appA3}), and we show the maximum ranges of the shift as a worst case estimate.
In the experiment, we do not spin-polarize the sample, and we  likely work with an equal population among all $m_F$ states.
As a conservative estimate, we use the full span due to the vector shift to estimate differential shifts between the measured tune-out detuning $\Delta_t^{88}$ for \Sr{88} and the tune-out detuning $\Delta_t^{87}$ for \Sr{87}.
By choosing to work with $\Delta_t^{88}$, we thus estimate a mismatch of $\pm\unit{23}{MHz}$, corresponding to a residual ground state polarizability of $\pm$\unit{0.05}{a.u.} for the AC Stark shift measurements presented in the main text.

\begingroup
\renewcommand{\arraystretch}{1.4}
\begin{table}
  \centering
  \begin{tabular*}{\columnwidth}{l@{\extracolsep{\fill}}ccc}
    \hline
    \hline
    $\alpha_g(\TPO, \omega)$  & $\delta\Delta_{t}=\Delta_{t}^{87}-\Delta_{t}^{88} $  \\
    \hline
    $\alpha_g^s(\TPO;\omega)$ & $+2\pi\times\unit{8}{MHz}$ \\
    $\alpha_g^v(\TPO;\omega)$ &  $-2\pi\times\unit{23}{MHz} \leq\delta\Delta_{t}\leq +2\pi\times\unit{23}{MHz}$\\
    $\alpha_g^t(\TPO;\omega)$  & $-2\pi\times\unit{2}{MHz} \leq\delta\Delta_{t}\leq +2\pi\times\unit{12}{MHz}$\\
    \hline
    \hline
  \end{tabular*}
  \caption{Comparison of the \Sr{87} and \Sr{88} tune-out detuning: $\Delta_{t}^{87}$ was numerically computed considering the contributions shown in the left column. For the vector and tensor contributions, the shift was calculated for the stretched $|m_{F}| = F$ states and we show the corresponding ranges. A beam ellipticity of 2\% was used for the vector shifts and we assumed perfect (maximum) polarization alignment for the tensor shifts.
  }
  \label{table_appendixA1}
\end{table}
\endgroup

Putting everything together, the scalar parts can be explicitly written as
\begin{equation}
  \begin{aligned}
    \alpha_g^{s}(\TPO;\omega) &= \frac{6\pi\epsilon_{0} c^3}{{\bar{\omega}_{\TPO}}^3 \tau_{\TPO}}\sum_{F_{j}}\left\{\frac{4}{15},\frac{5}{15},\frac{6}{15}\right\}\frac{\omega_{\TPO,F_{\TPO}}}{\omega^{2}_{\TPO,F_{\TPO}}-\omega^{2}}, \\ \alpha_g^{s}(\SPO;\omega) &\approx \frac{6\pi\epsilon_{0} c^3}{\tau_\SPO{}\omega_{\SPO}^2(\omega^{2}_{\SPO}-\omega^{2})}, \\
  \end{aligned}
\end{equation}
where we have neglected the hyperfine splitting in $\alpha_g^{s}(\SPO;\omega)$.
These approximations result in an additional shift of $\Delta_t$ by a few hundred kHz, which is more than two orders of magnitude smaller than our experimental precision.
As a consequence, we can express $\tau_{\SPO}$ directly in terms of the tune-out frequency
\begin{equation}
  \label{eq:appA14}
  \tau_{\SPO} = \frac{-6\pi\epsilon_{0} c^3}{[\alpha_g(v; \omega_t) + \alpha_g(c; \omega_t) + \alpha^{s}_g(\TPO, \omega_t)]\omega_{\SPO}^2(\omega^{2}_{\SPO}-\omega_t^{2})}
\end{equation}
For \Sr{88}, where the \TPO{} hyperfine splittings are absent, no approximations are necessary.
Thus, $\alpha^{s}_g(\TPO, \omega_t)$ reduces to
\begin{equation}
  \label{eq:appA14}
    \alpha_g^{s}(\TPO;\omega_t) = \frac{6\pi\epsilon_{0} c^3}{\tau_{\TPO}{\omega^2_{\TPO}}(\omega^{2}_{\TPO}-\omega_t^{2})}.
\end{equation}
To derive a fitting function to model the induced loss rate $\Gamma_{\text{mod}}\propto(\alpha^s_{g})^2$, we first need to express $g^{(0)}_{\TPO,F_\TPO}$ explicitly in terms of the laser detuning $\Delta = \omega - \bar{\omega}_{\TPO}$,
\begin{equation}
  \begin{aligned}
  g^{(0)}_{\TPO,F_\TPO}(\omega) &= \frac{1}{\hbar} \frac{\omega_{\TPO,F_\TPO}}{\omega_{\TPO,F_\TPO}^2 - \omega^2}\\
  &= \frac{1}{2\hbar} \bigg(\frac{1}{\omega_{\TPO,F_\TPO}-\omega}+\frac{1}{\omega_{\TPO,F_\TPO}+\omega}\bigg)\\
  & \approx \frac{1}{2\hbar}\bigg(\frac{1}{\Delta_{\TPO,F_{\TPO}}-\Delta}+\frac{1}{2\bar{\omega}_{\TPO}}\bigg),\\
  \end{aligned}
\end{equation}
where we use $\omega_{j,F_j}=\bar{\omega}_j+\Delta_{j,F_{j}}$ and $2\bar{\omega}_\TPO\gg\Delta+\Delta_{\TPO,F_{\TPO}}$. For the case of \Sr{88}, $\bar{\omega}_{j}$ is replaced by  $\omega_{j}$.
The last approximation on $g^{(0)}_{\TPO}$ shifts $\Delta_t$ by only several kHz.
Then, $\alpha_g^{s}(j;\omega)$ can be written as a function of the detuning as
\begin{equation}
  \label{eq:appA10}
  \begin{aligned}
  &\alpha_g^{s}(\TPO;\Delta) \approx \frac{1}{\tau_\TPO} \frac{3\pi\epsilon_0 c^3}{2\bar{\omega}^4_\TPO}\bigg[1-\\
  &\quad\qquad\qquad 2\sum_{F_\TPO} \left\{\frac{4}{15},\frac{5}{15},\frac{6}{15}\right\}\frac{\bar{\omega}_\TPO}{\Delta-\Delta_{\TPO,F_{\TPO}}}\bigg].
  \end{aligned}
\end{equation}
From this equation, we see that $\alpha_g^{s}(\TPO;\Delta)$ scales inversely proportional to detuning.
As shown in Fig.~\ref{fig:appendixA}(c), $\alpha_g^s(\SPO; \Delta)$ varies at the $10^{-3}$ level in the detuning range of our experimental data.
Thus, we treat $\alpha_g^{s}(\SPO;\Delta)$ as a constant $\alpha_g^{s}(\SPO;\Delta) =\alpha_g^{s}(\SPO;\Delta_t)$, which shifts $\Delta_t$ by less than a kHz.
Focusing on \Sr{88} where the hyperfine structure is absent, we arrive at the following expression when treating $\alpha_{\text{bg}}$ also as a constant in the vicinity of $\Delta_{t}$,
\begin{equation}
  \label{eq:appA20}
  \alpha_g(\Delta_{\TPO}) = \mathrm{const}\times \left(1-\frac{\Delta_{t}}{\Delta}\right).
\end{equation}
where we used the fact that $\alpha_g$ vanishes at the tune-out wavelength.
We use the function above to fit the induced loss rates, $\Gamma_{\text{mod}} \propto \alpha^{2}_g(\Delta_{\TPO})$.

\section{Fourier filtering}
\label{sec:fourier_filtering}

To suppress amplified spontaneous emission (ASE) that would induce systematic shifts in our measurements we took great care in filtering it.
Of particular importance is the amount of ASE near the \SSZ{}-\TPO{} transition where the polarizability diverges.
Even very low light levels near this transition lower the lifetime of atoms in the magic wavelength lattice and introduce an uncontrolled heating mechanism that is not easily accounted for.

To minimize the amount of light away from the carrier, we Fourier filter the tune-out laser beam with a grating as the dispersive element.
We expand the beam to a $1/e^2$ waist of \unit{2}{cm} and diffract it from a holographic grating with 2400~lines/mm, propagate it over $\sim$\unit{6}{m}, before finally coupling it into a single-mode polarization-maintaining fiber.

We determine the achieved resolution by measuring the transmission through the fiber while changing the input wavelength, at the same time keeping the grating angle fixed.
The Gaussian suppression line shape has a FWHM of \unit{13}{GHz}, giving a suppression of more than \unit{40}{dB} at the \SSZ{}-\TPO{} frequency when the input light is at the tune-out frequency.
Before filtering, we measure an ASE level of $-\unit{50}{dB}$ compared to the carrier with a spectrum analyzer with a \unit{0.05}{nm} resolution.
Together with the achieved Fourier filtering suppression of \unit{40}{dB} we conclude that the ASE level is reduced to below $-\unit{90}{dB}$ compared to the carrier.

\section{Loading excited state atoms in the tune-out lattice}
\label{sec:fourier_filtering}

To load excited state \Sr{87} atoms into the tune-out lattice the experimental procedure is as follows.
We
(i) load a sample of \Sr{87} $g$ atoms into the magic wavelength lattice;
(ii) after \unit{50}{ms}, we transfer $\sim$70\% of the atoms to $e$ using a 10-ms-long adiabatic-rapid-passage pulse;
(iii) after another \unit{2}{ms}, we diabatically switch on the tune-out lattice using \unit{64}{mW} of power;
(iv) after an additional \unit{1}{ms}, we ramp down the magic wavelength lattice over \unit{10}{ms} and retain $\sim$80\% of the $e$ atoms in the tune-out lattice;
(v) we hold the atoms in the tune-out lattice for a given time, diabatically turn it off, repump the atoms to $g$ over \unit{2}{ms}~\cite{snigirev19}, and take an absorption image to determine the final atom number.
In addition to matching the trapping frequencies of the tune-out and magic wavelength lattices, care was taken to achieve good mode matching between the two lattice beams. As a result the size and position of the area occupied by the trapped $e$ atoms in the tune-out and the magic lattice differs on the single micron level.
This is also supported by a measured moderate increase of the atom temperature to \unit{3}{\mu K} after the transfer into the tune-out lattice.


%